\documentclass[aps,preprint,epsfig,rotate]{revtex4}
\usepackage{graphicx}
\usepackage{bm}
\usepackage{epsfig}
\usepackage{pdflscape}
\usepackage{amsmath}

\newcommand{\be}{\begin{equation}}
\newcommand{\ee}{\end{equation}}
\newcommand{\bea}{\begin{eqnarray}}
\newcommand{\eea}{\end{eqnarray}}

\begin{document}

\title{Hylleraas-configuration-interaction analysis of the low-lying states in the three-electron Li  
atom and Be$^+$ ion}

\author{Mar\'{\i}a Bel\'{e}n Ruiz}
\email[E--mail address: ]{maria.belen.ruiz@fau.de}  

\author{Johannes T. Margraf}
\email[E--mail address: ]{johannes.margraf@fau.de}                                  

\affiliation{Department of Theoretical Chemistry\\
Friedrich-Alexander-University Erlangen-N\"urnberg,
Egerlandstra\ss e 3, 91058 Erlangen, Germany}

\author{Alexei M. Frolov}
\email[E--mail address: ]{afrolov@uwo.ca}

\affiliation{Department of Applied Mathematics\\
 University of Western Ontario, London, Ontario N6H 5B7, Canada}

\date{\today}

\begin{abstract}
The total energies of twenty eight bound S-, P-, D-, F-, G-, H-, and I-states 
in the three-electron Li atom and Be$^+$ ion, respectively, are determined with the 
use of the Configuration Interaction (CI) with Slater orbitals and L-S eigenfunctions,
and the Hylleraas-configuration-interaction (Hy-CI) methods. We discuss the construction 
and selection of the configurations in the wave functions, optimization of the orbital 
exponents and advanced computational techniques. Finally, we have developed an  
effective procedure which allows one to determine the energies of the excited states in
three-electron atoms and ions to high accuracy by using compact wave functions. For the
ground and low lying excited states our best accuracy with the Hy-CI method was $\approx
1 \cdot 10^{-6}$ a.u. and $1 \cdot 10^{-4}$ a.u. for other excited states. Analogous
accuracy of the CI method is substantially lower $\approx 1 \cdot 10^{-3}$ a.u. Many of 
the rotationally excited (bound) states in the three-electron Li atom and Be$^+$ ion have 
never been evaluated to such an accuracy.
\end{abstract}

\maketitle

\newpage

\section{Introduction}

Nowadays the Li atom has become, like the He atom years before, a system to
test quantum chemistry and high precision atomic physics \cite{Drake}. 
The non-relativistic wave functions of three-electron atoms and ions are of great interest in applications
related to highly accurate evaluations of the lowest-order relativistic and QED
corrections. At this moment we do not have any closed procedure which can be used to construct 
Dirac-type, manifestly Lorentz-invariant wave functions for two- and three-electron systems.  

As a consequence, in actual applications such few-electron wave functions are approximated by
the solutions of the non-relativistic Schr\"{o}dinger equation. All corrections are
evaluated with the use of the regular Rayleigh-Schr\"{o}dinger perturbation theory. Therefore, 
the non-relativistic wave functions of three-electrons atoms and ions are of paramount
importance. On the other hand, the accuracy of modern laser-based atomic 
experiments allows one to determine many transition lines (or transition energies) in 
three-electron atoms and ions to an accuracy which could not be expected even twenty years ago. 
To match these experimental results we need to increase (and very substantially) the accuracy 
of our current non-relativistic three-electron wave functions.

In the last few years the low lying states of the Li atom have been calculated to the accuracy
from a nanohartree to beyond a picohartree ($1\cdot 10^{-9}-10^{-12}$ a.u.)
\cite{Sims-Li,AdamP,AdamD,PKP,YD}. The corresponding wave functions usually contain many thousands
of basis functions (or configurations). Such sets of basis functions used in these approaches
include Hylleraas \cite{Lars}, Hylleraas-configuration-interaction \cite{Sims1971} three-electron functions (and their close
modifications). Recently, also the four-dimensional gaussoid functions of the relative coordinates
(see, e.g., \cite{KT}, \cite{Pos} and earlier references therein) has started to used again for
accurate calculations of the three-electron atomic systems. An alternative approach is based on the
construction of the compact wave functions, which are constructed by selecting the most contributing
basis functions (or configurations) and intensive optimization of the non-linear parameters
\cite{LK,King-Li1,Thakkar,Frolov-Li,King-Be}.
 
For the calculation of properties it would be desirable to have at hand all
energies and wave functions for ground and all excited states. In
addition, these states should be calculated with approximately the same
accuracy. Moreover, numerous excited states of all symmetry types (S, P, D,
F, G, H, I, $\ldots $) are usually needed, e.g. for the calculation of the probability of
ionization. No less importantly, the computational time should be
acceptable. The example of the Li atom can serve to test methods
and techniques developed for the calculations of properties, such as
excitation energies, transition probabilities, ionization energies, analysis
of optical spectra, energy levels in confinement conditions, nuclear
reactions and $\beta^{\pm}$-decay, etc, see, e.g., Ref. \cite{King-Li2}

In this work we employ the Hylleraas-configuration-interaction method
(Hy-CI) and the Configuration Interaction (CI) method with Slater orbitals
and L-S eigenfunctions to calculate a number of states of the Li atom and Be$^{+}$
ion which lie below their respective energy limits of electronic ionization.
The determination of non-S states with the Hy-CI wave function is easy, since 
the wave function retains the orbital picture.  
In the next sections we will discuss the procedures for selecting the energetically important
configurations, and optimizing the orbital exponents in
order to calculate accurate compact wave function expansions. Using this
method we have obtained several benchmark energies.

\section{The Hy-CI and CI methods}

The Hy-CI method was proposed by Sims and Hagstrom \cite
{Sims1971,Hy-CI1,Hy-CI2}. The advantage of the Hy-CI method with respect to
the other Hylleraas-type methods is that only up to one interelectronic
coordinate $r_{ij}$ per configuration is introduced into the wave function
and therefore, the method can in principle be applied to any atom.
Calculations with the use of Hy-CI wave functions for few-electron atoms (from He 
to B) and for the H$_2$ molecule were reported in Refs.  
\cite{Sims-He,Ruiz-He,Sims-Li,Busse98,Sims-Be,Ruiz-B,Sims-H2}. The CI wave
function with Slater orbitals and L-S eigenfunctions can be considered a
basic part of the Hy-CI wave function. In this work we start our
calculations with the CI wave functions. In this respect, we follow the same
method as Weiss and Bunge \cite{Weiss,Bunge} and use relatively small
basis sets. Recent extensive CI calculations with Slater orbitals on Be and
B atoms which are more accurate can be found in Refs. \cite{Bunge-Be,Bunge-B}%
. Both Hy-CI and CI wave functions can be summarized in the following
expression:

\begin{equation}
\Psi =\sum_{p=1}^NC_p\Phi _p,\qquad \Phi _p=\hat{O}(\hat{L}^2)\hat{\mathcal{A%
}}\psi _p\chi 
\end{equation}
The Hy-CI and CI wave functions are linear combinations of $N$ symmetry
adapted configurations $\Phi _p$ and the coefficients $C_p$, which are determined
variationally. In this work, the symmetry adapted configurations are
constructed 'a priori' so that they are eigenfunctions of the angular
momentum operator $\hat{L}^2$. Another possibility would be the posterior
projection of the configurations over the proper spatial space, as
indicated in Eq. (1) by the projection operator $\hat{O}(\hat{L}^2)$, where $\hat{\mathcal{A}%
}$ is the antisymmetrization operator and $\chi$ is the spin eigenfunction. 

In the case of the Li atom, it is sufficient to use only one
spin-function (formally a linear combination of the two possible spin
eigenfunctions would be necessary): 
\begin{equation}
\chi =\left[ (\alpha \beta -\beta \alpha )\alpha \right] \ 
\end{equation}
This is because the energetic contribution of the second spin eigenfunction has been
proven to be small (in the order of $1\times 10^{-9}$ a.u. \cite
{PP,YD}). Moreover, the Slater determinants produced by the second spin
eigenfunction $(2\alpha \alpha \beta -\beta \alpha \alpha -\alpha \beta
\alpha )$ (due to antisymmetry) are repeated when considering
the first spin eigenfunction $(\alpha \beta \alpha -\beta \alpha \alpha )$.
The spatial part of the basis functions consists of Hartree products of Slater
orbitals: 
\begin{equation}
\psi _p=r_{ij}^\nu \prod_{k=1}^n\phi _k(r_k,\theta _k,\varphi _k),
\end{equation}
where $\nu =0,1$ are employed for CI and Hy-CI wave functions, respectively.
Powers $\nu >1$ are effectively reduced to $\nu =0,1$, since all even and odd
powers of $r_{ij}$ can be expressed as a product of $r_{ij}$ times a
polynomial in $r_i,r_j$ and angular functions.

The basis functions $\phi _p$, are products of Slater orbitals. For the CI wave functions presented in this work, we use 
 $s$-, $p$-, $d$-, $f$-, $g$-, $h$- and $i$-Slater orbitals. In contrast, for the 
Hy-CI wave functions we use only $s$-, $p$-, $d$- and $f$-Slater orbitals.  
Higher angular orbitals are in practice only required to obtain 
an accuracy in the nanohartree regime ($1\cdot 10^{-9}$ 
a.u.) or higher with the Hy-CI method (see Ref. \cite{Sims-Li}). We use unnormalized complex
Slater orbitals, for which the exponents are adjustable parameters. These are defined
as:

\begin{equation}
\phi (\mathbf{r})=r^{n-1}e^{-\alpha r}Y_l^m(\theta ,\varphi ).
\end{equation}
The spherical harmonics with Condon and Shortley phase \cite[p. 52]{Condon}
are given by: 
\begin{equation}
Y_l^m(\theta ,\varphi )=(-1)^m\left[ \frac{2l+1}{4\pi }\frac{(l-m)!}{(l+m)!}%
\right] ^{1/2}P_l^m(\cos {\theta })e^{im\varphi },
\end{equation}
where $P_l^m(\cos {\theta })$ are the associated Legendre functions. The
spherical harmonics and associated Legendre functions used in this work
are written explicitly in \cite[p. 14]{Stevenson}.

The Hamiltonian in Hylleraas coordinates may be written in the infinite nuclear 
mass model \cite{Ruiz1,Barrois}

\begin{multline}
\hat{H} =-\frac 12\sum_{i=1}^n\frac{\partial ^2}{\partial r_i^2}%
-\sum_{i=1}^n\frac 1{r_i}\frac \partial {\partial r_i}-\sum_{i=1}^n\frac
Z{r_i}-\sum_{i<j}^n\frac{\partial ^2}{\partial r_{ij}^2}-\sum_{i<j}^n\frac
2{r_{ij}}\frac \partial {\partial r_{ij}}+\sum_{i<j}^n\frac 1{r_{ij}} \\
-\frac 12\sum_{i\neq j}^n\frac{r_i^2+r_{ij}^2-r_j^2}{r_ir_{ij}}\frac{%
\partial ^2}{\partial r_i\partial r_{ij}}-\frac 12\sum_{i\neq
j}^n\sum_{k>j}^n\frac{r_{ij}^2+r_{ik}^2-r_{jk}^2}{r_{ij}r_{ik}}\frac{%
\partial ^2}{\partial r_{ij}\partial r_{ik}} \\
-\frac 12\sum_{i=1}^n\frac 1{r_i^2}\frac{\partial ^2}{\partial \theta _i^2}%
-\frac 12\sum_{i=1}^n\frac 1{r_i^2\sin ^2{\theta _i}}\frac{\partial ^2}{%
\partial \varphi _i^2}-\frac 12\sum_{i=1}^n\frac{\cot {\theta _i}}{r_i^2}%
\frac \partial {\partial \theta _i} \\
-\sum_{i\neq j}^n\left( \frac{r_j}{r_ir_{ij}}\frac{\cos {\theta _j}}{\sin {%
\theta _i}}+\frac 12\cot {\theta _i}\frac{r_{ij}^2-r_i^2-r_j^2}{r_i^2r_{ij}}%
\right) \frac{\partial ^2}{\partial \theta _i\partial r_{ij}} \\
-\sum_{i\neq j}^n\frac{r_j}{r_ir_{ij}}\frac{\sin {\theta _j}}{\sin {\theta _i%
}}\sin {(\varphi _i-\varphi _j)}\frac{\partial ^2}{\partial \varphi
_i\partial r_{ij}}\ .
\end{multline}
The angular momentum operator can be extracted from Eq. (6):

\begin{equation}
\sum_{i=1}^n\frac 1{r_i^2}\hat{L}_i^2=-\frac 12\sum_{i=1}^n\frac 1{r_i^2}%
\frac{\partial ^2}{\partial \theta _i^2}-\frac 12\sum_{i=1}^n\frac
1{r_i^2\sin ^2{\theta _i}}\frac{\partial ^2}{\partial \varphi _i^2}-\frac
12\sum_{i=1}^n\frac{\cot {\theta _i}}{r_i^2}\frac \partial {\partial \theta
_i},
\end{equation}
and its eigenvalue equation is:

\begin{equation}
L_i^2\phi _i=l_i(l_i+1)\phi _i,
\end{equation}
with $l_i$ being the angular momentum quantum number of the orbital $\phi _i$. In the case
of Hy-CI wave functions, the variables $\partial ^2/(\partial r_{ij}\partial 
r_{ik})$ vanish. 

The kinetic energy operator has been separated into several radial and
angular parts. This operator has the advantage that, for the case of
three-electron kinetic integrals, the expansion of $r_{ij\text{ }}$into $%
r_{<} $ and $r_{>}$ is avoided, and therefore no three-electron auxiliary
integrals $W$ are required, see Ref. \cite{Ruiz3ek}. This fact saves not only calculations,
but also memory space. Only the easily computed two-electron auxiliary
integrals $V(n,m;\alpha ,\beta )$ are needed.

From the variational principle, one obtains the matrix eigenvalue problem:

\begin{equation}
(\mathbf{H}-E\mathbf{S})\mathbf{C}=\mathbf{0,}
\end{equation}
where the matrix elements are:

\begin{equation}
H_{kl}=\int \Phi _kH\Phi _ld\tau ,\text{ \qquad }S_{kl}=\int \Phi _k\Phi
_ld\tau .
\end{equation}

The integrals occurring in the Hy-CI calculations of three-electron systems, can
be divided into two- and three-electron integrals. The two-electron integrals are of the
types:

\begin{eqnarray}
&&\langle r_{12}\rangle ,\qquad \langle r_{12}^2\rangle ,\qquad \biggl< %
\frac 1{r_{12}}\biggr>,  \nonumber \\
&&\langle r_{12}\rangle \langle r_{34}\rangle ,\qquad \langle r_{12}\rangle \cdot  
\biggl<\frac 1{r_{34}}\biggr>,
\end{eqnarray}
where the notation $\langle r_{12}\rangle$ represents an integral, in which  
orbitals of electrons $1$ and $2$ are involved on the left and right-hand side, e.g.: 

\begin{equation}
\left\langle
\phi (\mathbf{r}_1) \phi (\mathbf{r}_2) r_{12} \phi (\mathbf{r}_1) \phi (%
\mathbf{r}_2) \right\rangle
\end{equation}
These two-electron integrals were evaluated
with the algorithms described in Ref. \cite{Ruiz2e}. 

The three-electron integrals are of the following types:
\begin{equation}
\biggl<r_{12}r_{13}\biggr>,\qquad \biggl<r_{12}^2r_{13}\biggr>,\qquad \biggl<%
\frac{r_{12}}{r_{13}}\biggr>,\qquad \biggl<\frac{r_{12}r_{13}}{r_{23}}\biggr>.
\end{equation}
The first three cases are evaluated by direct integration over one $r_{ij}$ and
the coordinates of one electron. They are thus reduced to a linear combination 
of two-electron integrals \cite{Ruiz3e}. 
For the so-called triangle integrals $\left\langle
r_{12}r_{13}/r_{23}\right\rangle $ we use a
very efficient subroutine by Sims and Hagstrom \cite{Sims3e}. Finally the
two- and three-electron kinetic energy integrals are evaluated using
the Hamiltonian of Eq. (6) \cite{Ruiz2e,Ruiz3ek}.

The integration of these three-electron integrals leads
to a limited linear combination of two-electron integrals. These can be
calculated very accurately in terms of two-electron auxiliary integrals $%
V(m,n;\alpha ,\beta )$, defined as:
\begin{equation}
V(m,n;\alpha ,\beta )=\int_0^\infty r_1^me^{-\alpha
r_1}dr_1\int_{r_1}^\infty r_2^ne^{-\beta r_2}dr_2\ ,
\end{equation}
The two-electron auxiliary integrals with positive indices $m,n$ are
in turn evaluated in terms of one-electron
auxiliary integrals $A(n,\alpha )$ \cite{Frolov-A}. 

In summary, only two-electron integrals, as in the CI method, and triangle
integrals have to be computed. This fact will be extremely helpful when
extending the application of the Hy-CI method to larger systems. In our code,
approximately the same amount of memory is required for CI and Hy-CI calculations. 
Note also that, if the Hy-CI method is applied to many-electron atoms/molecules,  
the highest order of required electron integrals is four.

To perform these computations, we have written a three-electron Hy-CI
computer program for three-electron systems in Fortran 90. The calculations were 
conducted with the use of quadruple precision arithmetics. The program has been 
thoroughly checked by comparing results of our numerical calculations with the
results by Sims and Hagstrom \cite{Sims-Li} and King \cite{King-Li1} for the
lithium atom. In these calculations we have obtained complete
agreement.

\section{Calculations}

\subsection{Construction and selection of the symmetry adapted configurations
}

The ground state configuration of the Li atom and Be$^{+}$ ion is $sss$ (i.e. $s(1)s(2)s(3)$). 
The further considered configurations for S-symmetry states (L=0) 
are, ordered by decreasing energetic contribution, $spp$, $pps$, $sdd$, $dds$, $sff$ and $ffs$.
The energetically important configurations for $L=0-6$ are listed in Table
I. The quantum number M=0 was chosen, because for this case a
smaller number of Slater determinants is required. We performed a systematical selection of the CI  
configurations according to their energy contribution. This was done by performing 
calculations on blocks constructed for all possible configurations. 
The eigenvalue equation was diagonalized upon each addition of a configuration. In this manner, the contribution  
of every single configuration and of each block of a given type to 
the total energy was evaluated. Configurations with an overall energy contribution below $1\cdot 10^{-8}$ a.u. 
were not considered.  

Usually the contribution of a configuration is larger, the smaller the sum of the $l$ quantum numbers 
of the employed orbitals $l_1+l_2+l_3$ is; i.e. 
the contribution of the configuration $ssp > ppp$ for a $P$-state. 
In cases such as the P states $spd$ and $ppp$,  where the sum of $l_i$ is equal, the two inner electrons in $ppp$ 
form a S-configuration. The resulting three electron configuration 
is $(^1S)p$ (a P-configuration), and contributes more than the $spd$ one. This is especially important in the case of 
F-, G-, H-, and I-states. Among the many possibilities to construct configurations of these symmetries,  
the energetically most important configurations were proven to be those with an inner S-shell and a single occupied 
orbital with the symmetry of the state under consideration, i.e. $(^1S)f$, $(^1S)g$, $(^1S)h$, and $(^1S)i$. The inner shell is described with a 
sum of configurations $(^1S)= ss+pp+dd+ff+gg+hh+ii$. In the CI calculations of S, P, and D states we employed s-, p-, d-, and f-orbitals (see Table I). 
In the CI calculations of the F, G, H, and I states we have used in addition g-, h-, and i-orbitals as shown in Table I. 
The energetic order determined for the CI calculations was kept for the Hy-CI calculations, where every CI configuration 
is multiplied by an interelectronic distance: 
Hy-CI = CI$\cdot\{1 + r_{12} + r_{13} + r_{23}\}$. 
    
Obviously, more types of configurations than the ones discussed here can be constructed for a given L quantum number. 
For instance, configurations like $psp$ could be considered, if the exponents $\alpha _1\ne \alpha _2$. However, we kept the orbital exponents in the K-shell equal, see Tables II  and
III. Therefore, the configuration $psp$ is equivalent to the configuration $spp$.  
Other possible higher energy configurations like $ppp$ for L=0, M=0 exist,
but were discarded due to their energetic contribution. Table I shows how the configurations used in this work were
constructed from $s$-, $p$-, $d$-, $f$-, $g$-, $h$-, and $i$-Slater orbitals. 

Finally, there are more possible 'degenerate L-eigenfunction' solutions with a larger number of
Slater determinants. Specifically, these are degenerate with respect to the quantum numbers L and M,
but with possible different energy contribution, i.e. non-degenerate with 
respect to the energy \cite{Bunge}. Although the
inclusion of various degenerate configurations has been shown to improve
the energy of the state, this contribution is very small. This is important
for very accurate CI calculations, as reported e.g. by Bunge  
\cite{Bunge-Be,Bunge-B}. In our work, we have concentrated on the energetically most
important CI configurations, in order to use them as the basis for Hy-CI
configurations (i.e. configurations multiplied by an interelectronic distance $%
r_{ij}$).

After selecting the types of configurations, we constructed complete
blocks of these configurations for a given basis set. For instance, for the
basis n=4 (i.e. [4s3p2d1f] or [1s2s3s4s2p3p3d4f]) in the $sss$ block
the following configurations were considered: $1s1s2s$, $1s2s2s$, $2s2s2s$, $1s1s3s$%
, $1s2s3s$, $2s2s3s$, \ldots , $4s4s4s$. Note that the configuration $1s1s1s$
has no physical meaning but displays a large energy contribution. Altogether, our CI
calculations can be considered 'selected' with respect to the type of
configuration, and 'full-CI' with respect to the orbitals basis set.

Another important aspect in CI and Hy-CI calculations is the symmetry
adaptation of the configurations. As mentioned above, the configurations are constructed 'a
priori' to be eigenfunctions of the angular momentum operator $%
\hat{L}^2$. In the sums of Table I, the configurations are formed by Slater
determinants. The determinants are pairwise symmetric (i.e. $sp_1p_{-1}$ and 
$sp_{-1}p_1$ in the $spp$ configuration) and lead to the same values of the
electronic integrals. Therefore, it is possible and desirable to consider
only one of the determinants and to deduce the result from the other.

In other words, the solution of the eigenvalue problem obtained when using  
reduced $1\times 1$ matrix
elements (where the integrals are added, configuration $%
sp_1p_{-1}+sp_{-1}p_1$) or when using explicit $2\times 2$ matrix elements of the
Slater determinants is the same. The symmetry adaptation is
computationally favorable, since the number of Slater determinants in the input is
smaller and the repeated computation of equal integrals is avoided. As can be seen in Table I, 
this procedure may be applied to all the constructed configurations.

The Hy-CI configuration blocks were constructed by including (1)
the corresponding CI block; (2) the CI block multiplied by the interelectronic 
coordinate $r_{12}$; (3) the CI block multiplied
by $r_{13}$; and (4) the CI block multiplied by $r_{23}$. Here, one has to take
into account possible symmetries between equivalent configurations. This
 can produce linear dependences which cause the calculation to break
down (due to linearly dependent equations in the eigenvalue problem). For example, $%
2s2s3s\cdot r_{13}$ is equal/equivalent to $2s2s3s\cdot r_{23}$.

In general, energetically important Hy-CI configurations must not be the
same as the corresponding CI ones, but usually this is the case. Therefore,
we constructed Hy-CI blocks of configurations based on the selected CI
ones. The number of configurations grows very fast when adding the three $%
r_{ij}$ factors. Therefore we filtered the configurations within a
block one by one, calculating the total energy $E_i$ everytime that a single configuration was added, 
and comparing it to the total energy without this configuration $E_{i-1}$. 

Again, if the difference of the energy was smaller than the   
energy criterion $|E_{i-1}-E_i|<1\cdot 10^{-8}$a.u, the new configuration was discarded. In
this manner, all configurations were checked, leading to a relatively compact Hy-CI wave function. 
Since the configuration selection process was carried out for every state,
the length of the final wave functions and the configurations included differ from
state to state and between Li and Be$^{+}$. This is natural, since we need different configurations to
describe different excited states.

\subsection{Optimization of the orbital exponents}

The orbital exponents were optimized for each atomic state of the Li
atom and Be$^{+}$ ion. A set of two exponents was used (one for the
K-shell and the other for the odd-electron in the L-shell), and kept equal
for all configurations. This technique accelerates the computations, while still
producing sufficiently accurate results for the calculation
of properties. It is clear that, for highly accurate energies beyond
microhartree-accuracy ($1\cdot 10^{-6}$ a.u.), more flexibility in the exponents is needed, as
shown in recent calculations on the lithium atom with extensive optimization 
\cite{Sims-Li,King-Li1,Frolov-Li,Thakkar} or in calculations with very large
wave functions and carefully chosen exponents \cite{YD,PKP,PP,Sims-Li}. 

The virial factor:
\begin{equation}
\chi = -\frac{\langle V \rangle}{\langle T \rangle}
\end{equation}
is used to check the quality of the wave function and guides the numerical
optimization of the exponents in the trial wave functions.  
In general, it is observed that the accuracy obtained in the virial factor, 
predicts approximately the number of the accurate decimal digits in the energy. For instance, the ground 
state energy of the Li atom has been calculated to -7.478 058 893 a.u. (6 decimal digits accurate) and its corresponding virial
factor is 2.000 000 954 (6 digits are zero), whereas the higher energy state $6 ^2S$ with energy -7.295 739 603 a.u. 
(3 decimal digits accurate) has a virial factor of 2.002 361.   

The optimization of two exponents at the same time, in the case of Li for all 
configurations, has the advantage that (being a global optimization) it is very fast, 
in contrast with the partial optimization of configurations one by one, 
which may take very long computational times.

The optimization of the orbital exponents was carried out via a parabolic procedure. 
Shortly, the orbital exponents are varied by a step size.
Three energy values are thus calculated and fitted to a parabola, and the minimum of
the parabola is calculated. Subsequently, this value is kept fixed and the same is
done for the next exponent. The step size is continually decreased by a
given factor as the cycles of exponent optimization are repeated. At every step the virial
factor is calculated. The optimization is performed until the
energy no longer improves, and the best virial
and energy values agree. The optimization program is completely
automatic and the exponents can be optimized for every state and
nuclear charge.

For the CI calculations, the orbital exponents were optimized until the same
energy minimum was obtained in two successive optimizations, starting with a basis of $n=4$. 
These exponents were then used in a CI calculation with the basis $n=5$, and
optimized again, and so on, up to the basis $n=7$. The optimized exponents of
the basis $n=7$ are reported in Table II. For Be$^{+}$ the same procedure
was repeated with the nuclear charge $Z=4$. The excited states were determined
by optimization of the orbital exponents for the second, third, $\ldots$  
eigenvalue. Note that in strictly variational methods, the successive
excited states are the roots of the eigenvalue problem. 
The exponent $\gamma$ of the singly occupied orbital gets smaller as the quantum number increases. 
We have obtained
energies that are about 1 millihartree accurate ($1\cdot 10^{-3}$ a.u.)
with respect to the non-relativistic values reported in the literature. 

The Hy-CI orbital exponents of the three lower states of every symmetry were optimized using a basis set $n=4$ of
about 400 configurations of all types considered. Subsequently, the exponents were
kept fixed for calculations with $n=5-8$ basis sets. The orbital
exponents for the higher excited states were optimized using few types of configurations (the 
energetically most important ones) and a larger basis set $n=8$.  
In Tables II and III the optimized exponents of the CI and Hy-CI wave functions are
given. Note that the CI exponents are in general larger than the Hy-CI ones.
This is in part because the Hy-CI wave functions employ a smaller
basis set.

The described method of optimization of the exponents is very successful for the 
determination of ground and low-lying excited states. For higher excited states, is 
not possible to obtain a good virial factor with this type of optimization. A larger orbital basis 
and more flexibility in the number of exponents would be needed.  

\section{Results}

We calculated S-, P-, D-, F-, G-, H- and I-symmetry states for the Li atom with the
CI method, using the symmetry adapted configurations shown in Table I. The CI calculations were carried 
out using double precision arithmetic (about 15 decimal digits accuracy on our workstations). In this 
manner, we determined the energy of seven S-states, six P-states, five D-states, four F-states, three 
G-states, two H-states and one I-state. Several of these states are reported for the first time. 

The total energies of the twenty eight states of the Li atom 
considered in this study are below the total energy of the ground state of the
${}^{\infty}$Li$^{+}$ ion, i.e. $E_{tr}$ $\approx$ -7.27991 34126 69305 96491 810(15) a.u.
 \cite{Frolov-Li+}. This total energy of the ground state in the two-electron 
${}^{\infty}$Li$^{+}$ ion is the natural threshold energy for an arbitrary bound 
state in the three-electron Li atom.

For the Be$^+$ ion we determined the total energies of the twenty eight bound states, 
including seven S-states, six P-states, five D-states, four F-states, three G-states, 
two H-states and one I-state. The results 
of our calculations can be found in Tables IV and V. In all these
calculations we applied the CI method. The computed energies are lower than the 
corresponding ionization energy of the Be$^+$ ion \cite{Ruiz-He}. The accuracy of the 
calculations is $\approx$ 1 millihartree ($1\cdot 10^{-3}$ a.u.). Note that the F-, G-, H- and I-states calculated with 
the CI method are reported here for the first time. 

For Hy-CI calculations, we employed the same blocks of configurations
as in the CI calculations, see Table I, and added blocks of these configurations
multiplied by one interelectronic coordinate at a time, i.e. CI$\cdot 
(1+r_{12}+r_{13}+r_{23}$). Details on the selection of configurations are 
given above. Hy-CI calculations up to the basis $n=6-8$ were performed. 

It is important to note that, in Hy-CI calculations it is usually not
necessary to use basis sets as large as in CI calculations, 
since the wave function expansion converges
faster to the exact solution. This is due to the explicit inclusion of the
interelectronic coordinate in the wave function. In contrast, in the CI
method the interelectronic coordinate is not explicitly considered, and its
effect is replaced by the use of high angular momentum orbitals. In short,
for Hy-CI calculations high angular momentum orbitals ($l\ge 3$) are not
required to achieve an accuracy in the microhartree regime ($1\cdot 10^{-6}$ a.u.), which is the purpose of this
paper. Note that highly accurate Hy-CI calculations can be afforded if using long wave functions expansions, 
see the benchmark energy values for the 6$^2S$ and 7$^2S$ states of Li atom \cite{Sims-Li}. 

The bound, rotationally excited F-, G-, H- and I-states have never been calculated with the use 
of the Hy-CI method due to the complexity of the related problems. Some recent developments, however,
make such calculations possible. For instance, in our computer program, the electronic integrals
are defined for every $l$ quantum number, but the kinetic energy integrals are currently restricted 
to $l\ge 2$, see Ref. \cite{Ruiz3ek}. The theoretical and computational implementation of higher 
quantum numbers is somewhat cumbersome, and will be reported elsewhere.

The non-relativistic total energies of the four and/or five lowest bound states of S-, P- and 
D-symmetry are now known to high accuracy, whereas other similar states have been determined to 
less accuracy, since in these cases we have used shorter trial wave functions (see Tables VI and VII). 

In the Hy-CI calculations of the Li atom, see Table VI, the maximum achieved accuracy is of a few 
microhartrees ($1 \cdot 10^{-6}$ a.u.) for the ground state and first S-, P-, and D-excited states. For higher 
excited states the accuracy is slightly lower. For highly excited bound states it is less 
than 1 millihartree ($1\cdot 10^{-3}$ a.u.), as far as values for these states were known. This is because
the use of a set of two exponents for all configurations is not as appropriate for 
highly excited states as for the ground and low-excited bound states. For instance, we determined the total
energy of the $8^2S$ ($E=$ -7.286 995 428 a.u.) state. By performing analogous calculations for the $4^2D$-state 
we obtained the total energy which is the best-to-date for this state $E =$ -7.311 211 047 a.u.  

For the Be$^+$ ion, the maximal accuracy is slightly better, which is directly related with the larger 
nuclear charge (see Table VII) and more compact electron wave function. For the ground state and low excited states of the Li atom  
we obtained an accuracy of few microhartrees ($1\cdot 10^{-6}$ a.u.), using less than 1000 configurations, whereas the best calculations 
in the literature use up to 14 000 configurations. 
 
The benchmarks obtained for Be$^+$ ion are: the $4^2P$ state ($E=$ -13.783 574 124 a.u.), 3$^2D$ state ($E=$ -13.878 041 021 a.u.), 
4$^2D$ state ($E=$ -13.780 663 883 a.u.), and 5$^2D$ state ($E=$ -13.735 537 780 a.u.). The newly calculated states are: 
$7 ^2S$ ($E=$ -13.699 224 475 a.u.), 8$^2S$ ($E=$ -13.687 885 004 a.u.), $7^2P$ ($E=$ -13.696 356 527 a.u.), 
$6^2D$ ($E=$ -13.710 204 495 a.u.) and $7^2D$ ($E=$ -13.695 419 936 a.u.).   
The dissociation threshold for the three-electron ${}^{\infty}$Be$^{+}$ ion is $\approx$ -13.65556 62384 23586 70207 810(15) a.u. \cite{Frolov-Li+}. 
This value coincides with the total energy of the ground $1^1S-$state of a Be$^{2+}$ ion with an infinitely heavy nucleus. The optical spectra of the
Li atom and Be$^{+}$ ions can be found in \cite{Strig}. The optical spectrum of the Li atom determined in this study is in a good agreement with the 
spectrum of the Li-atom shown in that work.  

We obtained this accuracy with less than 1 \% of the configurations used in the most highly accurate calculations 
reported. All calculated states are ordered by their energy and presented in Tables VIII and IX. The total energies of 
these states are below the corresponding threshold energy (or ionization energy) for the three-electron atomic systems considered 
here. 

\section{Conclusions and Perspectives}

We have determined the total energies of twenty eight bound states in the Li atom and Be$^{+}$ ion, respectively. The variational
wave functions of the S-, P-, D-, F-, G-, H-, and I-bound states in these three-electron atomic systems were constructed with the
use of the CI and Hy-CI methods. The procedure consisted in the appropriate selection of configurations and optimization of one set
of orbital exponents for every state. The total energies of the low-lying states are microhartree accurate ($1\cdot 10^{-6}$ a.u.),
while for excited states the accuracy is $\approx$ $1\cdot 10^{-4}-10^{-5}$ a.u. We have obtained several benchmarks and reported for
first time the energy of some highly excited states. These wave functions are convenient for the
calculation of properties. Consequently, these wave functions have been used
for the calculation of the transition probabilities during nuclear $\beta $%
-decay, where wave functions of very good quality are necessary to describe
the atomic effects during nuclear reactions \cite{our1,our2,our4}. Our future plan include
this systematic method of calculations to determine ground and excited states
of the following atoms and isoelectronic ions in the periodic table, such as
Be, B and C.

The results of our study are of great interest in various applications which includes different problems in Astrophysics
(e.g., to analyze the emission spectra of the hot Wolf-Rayet stars \cite{Allen}), Physics of Stellar and Laboratory plasmas,
Physics of Few-Body systems, etc. To the best of our knowledge, such extensive calculations of bound states in three-electron
have never been performed earlier with comparable accuracy. Analougous calculations of various rotationally and `vibrationally'
excited (bound) states in two-electron helium atom were conducted by Drake \cite{Drake2006}. It is clear that the total energies
reported by Drake \cite{Drake2006} are more accurate, but we consider a much more complicated case of bound states in
three-electron atoms and ions. Note also that the spectra of the two-electron atoms and ions include two independent series:
singlet and triplet, while for three-electron atoms/ions only doublet spin states belong to the actual discrete spectrum. All
quartet spin states of the three-electron atoms/ions are in the continuum, i.e. they are not truly bound states and any
interaction (such as spin-spin interactions) that breaks electron permutation symmetry will force the quadruplet states to
decay. Here we do not want to discuss the quadruple states in the three-electron atomic systems, since: (1) they are not truly
bound states, and (2) our method does not allow to analyze the properties of such states, which can be observed as quasi-bound
states at very special experimental conditions.

\section{Acknowledgments}

We are deeply indebted to Frederick King and James Sims for providing us with
numerical results of Hylleraas and Hylleraas-CI calculations, respectively, for the
lithium atom. These have proven useful in checking our computer program.
Very interesting discussions with James Sims and  Stanley Hagstrom
are greatly acknowledged.

\newpage



\begin{table}[tp]
\caption{List of the symmetry adapted configurations with quantum numbers
L=0-6 and M$_Z$=0 employed in the CI and Hy-CI calculations of the Li
atom and Be$^+$ ion. The notation $sss$ stands for $s(1)s(2)s(3)$. 
The Hy-CI configurations are obtained from the CI ones by
multiplying them by the factor $R=\{1 + r_{12} + r_{13} + r_{23}$\}. 
Normalization factors are omitted.}
\begin{center}
\scalebox{0.80}{
\begin{tabular}{ccc}
\hline\hline
L \qquad & Confs. & Construction \\ \hline
0 & $sss$ & $sss$ \\ 
0 & $spp$ & $sp_0p_0 - sp_1p_{-1} - sp_{-1}p_1$ \\ 
0 & $pps$ & $p_0p_0s - p_1p_{-1}s - p_{-1}p_1s$ \\ 
0 & $sdd$ & $sd_0d_0 - sd_1d_{-1} - sd_{-1}d_1 + sd_2d_{-2} + sd_{-2}d_2$ \\ 
0 & $dds$ & $d_0d_0s - d_1d_{-1}s - d_{-1}d_1s + d_2d_{-2}s + d_{-2}d_2s$ \\ 
0 & $sff$ & $sf_0f_0 - sf_1f_{-1} - sf_{-1}f_1 + sf_2f_{-2} + sf_{-2}f_2 -
sf_3f_{-3} - sf_{-3}f_3$ \\ 
0 & $ffs$ & $f_0f_0s - f_1f_{-1}s - f_{-1}f_1s + f_2f_{-2}s + f_{-2}f_2s -
f_3f_{-3}s - f_{-3}f_3s$ \\ 
1 & $ssp$ & $ssp_0$ \\ 
1 & $sps$ & $sp_0s$ \\ 
1 & $ppp$ & $p_0p_0p_0 - p_1p_{-1}p_0 - p_{-1}p_1p_0$ \\ 
1 & $ddp$ & $d_0d_0p_0 - d_1d_{-1}p_0 - d_{-1}d_1p_0 + d_2d_{-2}p_0 +
d_{-2}d_2p_0$ \\ 
1 & $pdd$ & $p_0d_0d_0 - p_0d_1d_{-1} - p_0d_{-1}d_1 + p_0d_2d_{-2} +
p_0d_{-2}d_{2}$ \\ 
1 & $spd$ & $sp_0d_0 - sp_1d_{-1} - sp_{-1}d_1 $ \\ 
1 & $pds$ & $p_0d_0s - p_1d_{-1}s - p_{-1}d_1s $ \\ 
1 & $sdp$ & $sd_0p_0 - sd_1p_{-1} - sd_{-1}p_1 $ \\ 
2 & $ssd$ & $ssd_0$ \\ 
2 & $sds$ & $sd_0s$ \\ 
2 & $spp$ & $sp_0p_0 + sp_1p_{-1} + sp_{-1}p_1$ \\ 
2 & $pps$ & $p_0p_0s + p_1p_{-1}s + p_{-1}p_1s$ \\ 
2 & $ppd$ & $p_0p_0d_0 - p_1p_{-1}d_0 - p_{-1}p_1d_0$ \\ 
2 & $ddd$ & $d_0d_0d_0 - d_1d_{-1}d_0 -d_{-1}d_1d_0 + d_2d_{-2}d_0 +
d_{-2}d_2d_0$ \\ 
2 & $spf$ & $sp_0f_0 - sp_1f_{-1} - sp_{-1}f_1$ \\ 
2 & $pfs$ & $p_0f_0s - p_1f_{-1}s - p_{-1}f_1s$ \\ 
2 & $sfp$ & $sf_0p_0 - sf_1p_{-1} - sf_{-1}p_1$ \\ 
3 & $ssf$ & $ssf_0$ \\ 
3 & $sfs$ & $sf_0s$ \\ 
3 & $ppf$ & $p_0p_0f_0 - p_1p_{-1}f_0 - p_{-1}p_1f_0$ \\ 
3 & $ddf$ & $d_0d_0f_0 - d_1d_{-1}f_0 - d_{-1}d_1f_0 + d_2d_{-2}f_0 +
d_{-2}d_2f_0$ \\ 
3 & $fff$ & $f_0f_0f_0 - f_1f_{-1}f_0 - f_{-1}f_1f_0 + f_2f_{-2}f_0 +
f_{-2}f_2f_0 - f_3f_{-3}f_0 - f_{-3}f_3f_0$ \\ 
\hline\hline
\end{tabular}
}
\end{center}
\end{table}

\newpage

\begin{turnpage}
\begin{table}[tp]
{Continuation TABLE I.} 
\begin{center}
\scalebox{0.80}{
\begin{tabular}{ccc}
\hline\hline
L \qquad & Confs. & Construction \\ 
\hline
3 & $ggf$ & $g_0g_0f_0 - g_1g_{-1}f_0 - g_{-1}g_1f_0 + g_2g_{-2}f_0 + g_{-2}g_2f_0 - g_3g_{-3}f_0 - g_{-3}g_3f_0 + g_4g_{-4}f_0 + g_{-4}g_4f_0$ \\
3 & $hhf$ & $h_0h_0f_0 - h_1h_{-1}f_0 - h_{-1}h_1f_0 + h_2h_{-2}f_0 + h_{-2}h_2f_0 - h_3h_{-3}f_0 - h_{-3}h_3f_0 + h_4h_{-4}f_0 + h_{-4}h_4f_0 - h_5h_{-5}f_0 - h_{-5}h_5f_0$ \\
4 & $ssg$ & $ssg_0$ \\
4 & $sgs$ & $sg_0s$ \\
4 & $ppg$ & $p_0p_0g_0 - p_1p_{-1}g_0 - p_{-1}p_1g_0$ \\
4 & $ddg$ & $d_0d_0g_0 - d_1d_{-1}g_0 - d_{-1}d_1g_0 + d_2d_{-2}g_0 + d_{-2}d_2g_0$ \\
4 & $ffg$ & $f_0f_0g_0 - f_1f_{-1}g_0 - f_{-1}f_1g_0 + f_2f_{-2}g_0 + f_{-2}f_2g_0 - f_3f_{-3}g_0 - f_{-3}f_3g_0$ \\                 
4 & $ggg$ & $g_0g_0g_0 - g_1g_{-1}g_0 - g_{-1}g_1g_0 + g_2g_{-2}g_0 + g_{-2}g_2g_0 - g_3g_{-3}g_0 - g_{-3}g_3g_0 + g_4g_{-4}g_0 + g_{-4}g_4g_0$ \\ 
4 & $hhg$ & $h_0h_0g_0 - h_1h_{-1}g_0 - h_{-1}h_1g_0 + h_2h_{-2}g_0 + h_{-2}h_2g_0 - h_3h_{-3}g_0 - h_{-3}h_3g_0 + h_4h_{-4}g_0 + h_{-4}h_4g_0 - h_5h_{-5}g_0 - h_{-5}h_5g_0$ \\
5 & $ssh$ & $ssh_0$ \\
5 & $shs$ & $sh_0s$ \\
5 & $pph$ & $p_0p_0h_0 - p_1p_{-1}h_0 - p_{-1}p_1h_0$ \\
5 & $ddh$ & $d_0d_0h_0 - d_1d_{-1}h_0 - d_{-1}d_1h_0 + d_2d_{-2}h_0 + d_{-2}d_2h_0$ \\
5 & $ffh$ & $f_0f_0h_0 - f_1f_{-1}h_0 - f_{-1}f_1h_0 + f_2f_{-2}h_0 + f_{-2}f_2h_0 - f_3f_{-3}h_0 - f_{-3}f_3h_0$ \\                 
5 & $ggh$ & $g_0g_0h_0 - g_1g_{-1}h_0 - g_{-1}g_1h_0 + g_2g_{-2}h_0 + g_{-2}g_2h_0 - g_3g_{-3}h_0 - g_{-3}g_3h_0 + g_4g_{-4}h_0 + g_{-4}g_4h_0$ \\             
5 & $hhh$ & $h_0h_0h_0 - h_1h_{-1}h_0 - h_{-1}h_1h_0 + h_2h_{-2}h_0 + h_{-2}h_2h_0 - h_3h_{-3}h_0 - h_{-3}h_3h_0 + h_4h_{-4}h_0 + h_{-4}h_4h_0 - h_5h_{-5}h_0 - h_{-5}h_5h_0$ \\
6 & $ssi$ & $ssi_0$ \\
6 & $sis$ & $si_0s$ \\
6 & $ppi$ & $p_0p_0i_0 - p_1p_{-1}i_0 - p_{-1}p_1i_0$ \\
6 & $ddi$ & $d_0d_0i_0 - d_1d_{-1}i_0 - d_{-1}d_1i_0 + d_2d_{-2}i_0 + d_{-2}d_2i_0$ \\
6 & $ffi$ & $f_0f_0i_0 - f_1f_{-1}i_0 - f_{-1}f_1i_0 + f_2f_{-2}i_0 + f_{-2}f_2i_0 - f_3f_{-3}i_0 - f_{-3}f_3i_0$ \\
6 & $ggi$ & $g_0g_0i_0 - g_1g_{-1}i_0 - g_{-1}g_1i_0 + g_2g_{-2}i_0 + g_{-2}g_2i_0 - g_3g_{-3}i_0 - g_{-3}g_3i_0 + g_4g_{-4}i_0 + g_{-4}g_4i_0$ \\
6 & $hhi$ & $h_0h_0i_0 - h_1h_{-1}i_0 - h_{-1}h_1i_0 + h_2h_{-2}i_0 + h_{-2}h_2i_0 - h_3h_{-3}i_0 - h_{-3}h_3i_0 + h_4h_{-4}i_0 + h_{-4}h_4i_0 - h_5h_{-5}i_0 - h_{-5}h_5i_0$ \\
6 & $iii$ & $i_0i_0i_0 - i_1i_{-1}i_0 - i_{-1}i_1i_0 + i_2i_{-2}i_0 + i_{-2}i_2i_0 - i_3i_{-3}i_0 - i_{-3}i_3i_0 + i_4i_{-4}i_0 + i_{-4}i_4i_0 - i_5i_{-5}i_0 - i_{-5}i_5i_0 
+ i_6i_{-6}i_0 + i_{-6}i_6i_0$ \\
\hline\hline
\end{tabular}
}
\end{center}
\end{table}
\end{turnpage}


\begin{table}[tp]
\caption{Orbital exponents used in the CI and Hy-CI calculations of the Li atom of Tables IV and VI. 
The shells are doubly occupied $\alpha=\beta$ and $\gamma$ is the 
exponent of the single occupied orbital. The virial factor has been obtained during the optimization of orbital exponents.}
\begin{center}
\scalebox{0.85}{
\begin{tabular}{rccccccc}
\hline\hline
No. \quad & State \quad & \quad $\alpha_{\mathrm{CI}}=\beta_{\mathrm{CI}}$ \qquad & \qquad $\gamma_{\mathrm{CI}}$
\qquad & \qquad virial$_{\mathrm{CI}}$ \quad & \quad $\alpha_{\mathrm{Hy-CI}}=\beta_{\mathrm{Hy-CI}}$ 
\quad & \quad $\gamma_{\mathrm{Hy-CI}}$ \quad & \quad virial$_{\mathrm{Hy-CI}}$ \\ 
\hline
1 & 2$^2S$ & 4.644060 & 1.107868 & 2.000000 & 2.994250 & 0.839625 & 2.000000 \\ 
3 & 3$^2S$ & 4.698079 & 0.561144 & 2.000000 & 3.550050 & 0.438800 & 2.000028 \\ 
6 & 4$^2S$ & 4.605560 & 0.359164 & 2.000196 & 3.241342 & 0.304425 & 2.000483 \\ 
10& 5$^2S$ & 4.639431 & 0.258794 & 2.001368 & 3.906990 & 0.235023 & 2.000534 \\         
15& 6$^2S$ & 4.602371 & 0.191464 & 2.003141 & 4.602371 & 0.191464 & 2.002362 \\            
24& 7$^2S$ & 4.696442 & 0.154444 & 2.002970 & 3.384262 & 0.134132 & 2.003046 \\                                                      
28& 8$^2S$ & 5.039868 & 0.094174 & 2.003949 & 2.857940 & 0.134896 & 2.003993 \\                                            
\hline
2 & 2$^2P$ & 4.451592 & 0.827973 & 2.000016 & 3.140842 & 0.725092 & 2.000000 \\ 
4 & 3$^2P$ & 4.507292 & 0.504441 & 2.000053 & 3.520217 & 0.370508 & 2.000087 \\ 
7 & 4$^2P$ & 4.486842 & 0.320982 & 2.000175 & 3.523842 & 0.255675 & 2.001194 \\ 
13& 5$^2P$ & 4.577346 & 0.230959 & 2.000776 & 3.844842 & 0.204309 & 2.001203 \\ 
19& 6$^2P$ & 4.513159 & 0.169635 & 2.003267 & 3.426592 & 0.170633 & 2.001447 \\ 
26& 7$^2P$ & 4.581002 & 0.144740 & 2.004915 & 3.971050 & 0.144667 & 2.003926 \\ 
\hline
5 & 3$^2D$ & 4.512037 & 0.459812 & 2.000000 & 3.359717 & 0.347508 & 2.000005 \\ 
8 & 4$^2D$ & 4.483217 & 0.247508 & 2.000012 & 3.496259 & 0.253341 & 1.999654 \\ 
14& 5$^2D$ & 4.483288 & 0.200160 & 2.000073 & 3.841288 & 0.200160 & 2.000234 \\           
20& 6$^2D$ & 4.539008 & 0.166704 & 2.000433 & 4.960208 & 0.150343 & 2.003830 \\ 
27& 7$^2D$ & 4.492642 & 0.143125 & 2.001133 & 4.492642 & 0.143125 & 2.000086 \\
\hline
9 & 4$^2F$ & 4.630029 & 0.299372 & 2.000027 &          &          &          \\ 
12& 5$^2F$ & 4.645336 & 0.186989 & 2.000041 &          &          &          \\ 
18& 6$^2F$ & 4.669615 & 0.174730 & 2.000061 &          &          &          \\ 
25& 7$^2F$ & 4.763930 & 0.141984 & 2.000367 &          &          &          \\ 
\hline
11& 5$^2G$ & 5.165095 & 0.234090 & 1.999983 &          &          &          \\
17& 6$^2G$ & 5.127263 & 0.207967 & 1.999997 &          &          &          \\
26& 7$^2G$ & 5.076262 & 0.142897 & 2.000050 &          &          &          \\
\hline
16& 6$^2H$ & 5.077561 & 0.181168 & 2.000011 &          &          &          \\ 
22& 7$^2H$ & 5.077536 & 0.141189 & 2.000021 &          &          &          \\
\hline
21& 7$^2I$ & 5.077551 & 0.144224 & 2.000014 &          &          &          \\
\hline\hline
\end{tabular}
}
\end{center}
\end{table}

\newpage


\begin{table}[tp]
\caption{Orbital exponents used in the CI and Hy-CI calculations of the Be$^ +$ ion of Tables V and VII. The shells are doubly occupied $\alpha=\beta$ and $\gamma$ is the 
exponent of the single occupied orbital. The virial factor has been obtained during the optimization of orbital exponents.}
\begin{center}
\scalebox{0.85}{
\begin{tabular}{rccccccc}
\hline\hline
No. \quad & State \quad & \quad $\alpha_{\mathrm{CI}}=\beta_{\mathrm{CI}}$ \qquad & \qquad $\gamma_{\mathrm{CI}}$
\qquad & \qquad virial$_{\mathrm{CI}}$ \quad & \quad $\alpha_{\mathrm{Hy-CI}}=\beta_{\mathrm{Hy-CI}}$ 
\quad & \quad $\gamma_{\mathrm{Hy-CI}}$ \quad & \quad virial$_{\mathrm{Hy-CI}}$ \\
\hline
1  & 2$^2S$ & 6.345407 & 1.950188 & 2.000000 & 4.173235 & 1.406372 & 2.000000 \\ 
3  & 3$^2S$ & 6.307744 & 1.019290 & 2.000017 & 4.838717 & 0.876476 & 2.000009 \\ 
6  & 4$^2S$ & 6.280890 & 0.665840 & 2.000087 & 4.766466 & 0.585863 & 2.000153 \\ 
10 & 5$^2S$ & 6.327154 & 0.479516 & 2.001115 & 5.331658 & 0.442445 & 2.001495 \\ 
15 & 6$^2S$ & 6.304912 & 0.364545 & 2.003618 & 5.150658 & 0.401445 & 2.000714 \\  
21 & 7$^2S$ & 6.341861 & 0.304815 & 2.004436 & 4.475332 & 0.287333 & 2.002761 \\ 
28 & 8$^2S$ & 6.397917 & 0.248083 & 2.005359 & 4.201068 & 0.260587 & 2.003766 \\ 
\hline                                                         
2  & 2$^2P$ & 6.141069 & 1.760345 & 2.000000 & 4.746625 & 1.321000 & 2.000002 \\ 
4  & 3$^2P$ & 6.148030 & 0.964810 & 2.000031 & 4.837058 & 0.712777 & 2.000045 \\ 
7  & 4$^2P$ & 6.158844 & 0.631221 & 2.000183 & 4.800125 & 0.516000 & 2.000749 \\ 
11 & 5$^2P$ & 6.168223 & 0.465315 & 2.001420 & 5.069200 & 0.532429 & 2.000760 \\ 
19 & 6$^2P$ & 6.189834 & 0.344862 & 2.006738 & 3.951567 & 0.339598 & 2.007193 \\ 
26 & 7$^2P$ & 6.222913 & 0.290957 & 2.010724 & 4.743317 & 0.289533 & 2.008245 \\
\hline
5  & 3$^2D$ & 6.132817 & 0.867614 & 2.000005 & 4.804284 & 0.670348 & 2.000004 \\ 
9  & 4$^2D$ & 6.159980 & 0.591028 & 2.000005 & 4.750784 & 0.588652 & 2.000012 \\ 
14 & 5$^2D$ & 6.134907 & 0.434209 & 2.000047 & 5.720102 & 0.381441 & 2.000363 \\ 
20 & 6$^2D$ & 6.484197 & 0.337802 & 2.000781 & 5.539008 & 0.443304 & 2.000114 \\     
27 & 7$^2D$ & 6.157877 & 0.286801 & 2.002368 & 4.849877 & 0.286801 & 2.001541 \\        
\hline
8  & 4$^2F$ & 6.440871 & 0.589239 & 1.999989 &          &          &          \\ 
13 & 5$^2F$ & 6.446382 & 0.372527 & 2.000019 &          &          &          \\ 
18 & 6$^2F$ & 6.352375 & 0.344411 & 2.000119 &          &          &          \\ 
25 & 7$^2F$ & 6.419948 & 0.285310 & 2.000794 &          &          &          \\ 
\hline
12 & 5$^2G$ & 6.844762 & 0.442011 & 2.000017 &          &          &           \\
17 & 6$^2G$ & 6.896092 & 0.360327 & 2.000012 &          &          &          \\
24 & 7$^2G$ & 7.030220 & 0.289953 & 2.000041 &          &          &          \\
\hline
16 & 6$^2H$ & 6.923097 & 0.337191 & 2.000007 &          &          &          \\
23 & 7$^2H$ & 6.902410 & 0.283624 & 2.000024 &          &          &          \\
\hline
22 & 7$^2I$ & 6.977235 & 0.288487 & 2.000000 &          &          &          \\
\hline\hline
\end{tabular}
}
\end{center}
\end{table}

\newpage


\begin{table}
\begin{center}
\caption{Convergence of Full-CI (L-S) calculations on the ground and excited states of the  Li atom with respect to the basis set.  
The basis sets are constructed with Slater orbitals, see Table I. The optimized orbital 
exponents for the largest basis are given in Table II. No. is the ordering number of the state. N is the number of symmetry adapted configurations (Table I). 
All energies are given in a.u., while Diff. are the energy differences between the present and reference 
energies in microhratrees ($1\cdot 10^{-6}$ a.u.).}
\scalebox{0.75}{%
\begin{tabular}{rcccccrlcr}
\hline\hline
No. \quad & State & \qquad  N \qquad &  n=6  &  \qquad N \qquad  &   n=7      & \qquad  N  \qquad  & \qquad \qquad Ref. Energy  & Ref.  & \qquad Diff. \\
\hline
1 & 2$^2S$ &  596 & -7.476 817 &  991 & -7.477 192 & 34020 & $\;$-7.478 060 323 910 146 894 & \cite{YD}    &  868.7 \\
3 & 3$^2S$ &  596 & -7.352 980 &  991 & -7.353 249 & 34020 & $\;$-7.354 098 421 444 364 045 & \cite{YD}    &  849.7 \\
6 & 4$^2S$ &  596 & -7.317 410 &  991 & -7.317 679 & 34020 & $\;$-7.318 530 845 998 906 901 & \cite{YD}    &  851.9 \\
10& 5$^2S$ &  596 & -7.302 342 &  991 & -7.302 682 & 34020 & $\;$-7.303 551 579 226 734 650 & \cite{YD}    &  870.0 \\
15& 6$^2S$ &  596 & -7.294 676 &  991 & -7.294 935 & 34020 & $\;$-7.295 859 510 844 131 039 & \cite{YD}    &  924.3 \\
24& 7$^2S$ &      &            &  991 & -7.289 596 & 17072 & $\;$-7.291 392 273 116         & \cite{Sims-Li} & 1796.3  \\ 
  &        &      &            &      &   n=8$^a$  &       &                              &                &        \\                                
28& 8$^2S$ &      &            &  508 & -7.285 695 &       &                              &                &        \\
\hline
2 & 2$^2P$ &  849 & -7.408 437 & 1430 & -7.408 619 & 32200 & $\;$-7.410 156 532 652 370     & \cite{YD}    & 1537.8 \\
4 & 3$^2P$ &  849 & -7.335 436 & 1430 & -7.335 658 &  7000 & $\;$-7.337 151 707 93          & \cite{AdamP} & 1493.5 \\ 
7 & 4$^2P$ &  849 & -7.310 200 & 1430 & -7.310 383 &  7000 & $\;$-7.311 889 059 38          & \cite{AdamP} & 1506.1 \\     
13& 5$^2P$ &  849 & -7.298 615 & 1430 & -7.298 802 &  7000 & $\;$-7.300 288 164 88          & \cite{AdamP} & 1486.0 \\ 
19& 6$^2P$ &  849 & -7.292 380 & 1430 & -7.292 545 &  7000 & $\;$-7.294 020 052 93          & \cite{AdamP} & 1475.3 \\
26& 7$^2P$ &      &            & 1430 & -7.288 749 &  7000 & $\;$-7.290 254 908 09          & \cite{AdamP} & 1506.0 \\ 
\hline
5 & 3$^2D$ &  646 & -7.333 935 & 1056 & -7.334 100 & 32760 & $\;$-7.335 523 543 524 685    & \cite{YD}     & 1423.5 \\
8 & 4$^2D$ &  646 & -7.309 598 & 1056 & -7.309 761 &  4000 & $\;$-7.311 189 578 43         & \cite{AdamD}  & 1428.1 \\
14& 5$^2D$ &  646 & -7.298 340 & 1056 & -7.298 502 &  4000 & $\;$-7.299 927 555 94         & \cite{AdamD}  & 1425.6 \\
20& 6$^2D$ &  646 & -7.292 225 & 1056 & -7.292 387 &  4000 & $\;$-7.293 810 713 64         & \cite{AdamD}  & 1423.5 \\
27& 7$^2D$ &      &            & 1056 & -7.288 700 &  4000 & $\;$-7.290 122 856 24         & \cite{AdamD}  & 1422.2 \\
\hline
9 & 4$^2F$ &  286 & -7.310 288 &  532 & -7.310 610 &       & $\;$-7.311 168 7           & \cite{King-Li1}  &  559.1 \\
12& 5$^2F$ &  286 & -7.298 989 &  532 & -7.299 340 &       & $\;$-7.299 917 1           & \cite{King-Li1}  &  576.8 \\
18& 6$^2F$ &  286 & -7.292 769 &  532 & -7.293 211 &       &                            &            &        \\
25& 7$^2F$ &      &            &  532 & -7.289 401 &       &                            &            &        \\
\hline
  &        &      &    n=7     &      &    n=8     &       &                            &            &        \\
11& 5$^2G$ & 395  & -7.299 248 & 694  & -7.299 430 &       &                            &            &        \\ 
17& 6$^2G$ & 395  & -7.293 125 & 694  & -7.293 294 &       &                            &            &        \\
26& 7$^2G$ & 395  & -7.289 383 & 694  & -7.289 605 &       &                            &            &        \\
\hline
16& 6$^2H$ & 272  & -7.293 138 & 519  & -7.293 320 &       &                            &            &        \\
22& 7$^2H$ & 272  & -7.289 435 & 519  & -7.289 625 &       &                            &            &        \\ 
\hline
21& 7$^2I$ &      &            & 350  & -7.289 638 &       &                            &            &        \\
\hline\hline
\end{tabular}}
\end{center}
\footnotetext[1]{For the calculation of the 8$^2S$, 7$^2G$, 7$^2H$ and 7$^2I$ states larger basis sets including n=8 orbitals are needed.}  
\end{table}

\newpage


\begin{table}[tp]
\begin{center}
\caption{Convergence of Full-CI (L-S) calculations on the ground and excited states of the Be$^+$ ion with respect to the basis set.  
The basis set and symmetry adapted configurations used are the same than for the Li atom calculations of Table IV. 
All energies are given in a.u., while Diff. are the energy differences between 
the present and reference energies in microhartrees ($1\cdot 10^{-6}$ a.u.).}
\scalebox{0.75}{%
\begin{tabular}{rcccccrlcr}
\hline\hline
No. \quad & State & \qquad  N \qquad &  n=6  &  \qquad N \qquad  &   n=7      & \qquad  N  \qquad  & \qquad \qquad Ref. Energy  & Ref.  & \qquad Diff. \\
\hline
1  & 2$^2S$ & 596 & -14.323 468 &  991 & -14.323 769 & 13944 & $\;$-14.324 763 176 790 43(22) & \cite{PKP}      &  994.6   \\
3  & 3$^2S$ & 596 & -13.921 529 &  991 & -13.921 830 &~10000 & $\;$-13.922 789 268 544 2      & \cite{PP}       &  959.2   \\
6  & 4$^2S$ & 596 & -13.797 444 &  991 & -13.797 754 &  1888 & $\;$-13.798 716 609 2          & \cite{Adam-Be+} &  962.8   \\
10 & 5$^2S$ & 596 & -13.743 267 &  991 & -13.743 655 &  1091 & $\;$-13.744 631 82             & \cite{King-Be+} &  977.0   \\ 
15 & 6$^2S$ & 596 & -13.714 814 &  991 & -13.715 222 &  2058 & $\;$-13.716 286 24             & \cite{King-Be+} & 1064.1   \\
21 & 7$^2S$ & 596 & -13.689 753 &  991 & -13.697 421 &       &                                &                 &          \\
   &        &     &             &      &    n=8      &       &                                &                 &          \\   
28 & 8$^2S$ &     &             &  991 & -13.684 764 &       &                                &                 &          \\
\hline
2  & 2$^2P$ & 849 & -14.177 210 & 1430 & -14.177 409 &~10000 & $\;$-14.179 333 293 342 7      & \cite{PP}       & 1924.4   \\
4  & 3$^2P$ & 849 & -13.883 174 & 1430 & -13.883 425 &       & $\;$-13.885 15                 & \cite{Woznicki} & 1725.3   \\
7  & 4$^2P$ & 849 & -13.781 745 & 1430 & -13.781 975 & 1021  & $\;$-13.783 518 3              & \cite{WZC1}     & 1543.5   \\
11 & 5$^2P$ & 849 & -13.735 200 & 1430 & -13.735 466 &       & $\;$-13.737 18                 & \cite{Woznicki} & 1714.1   \\
19 & 6$^2P$ & 849 & -13.710 140 & 1430 & -13.710 331 &       & $\;$-13.712 06                 & \cite{Woznicki} & 1729.2   \\
26 & 7$^2P$ &     &             & 1430 & -13.695 228 &       &                            &                 &          \\
\hline
5  & 3$^2D$ & 646 & -13.876 261 & 1056 & -13.876 447 &   841 & $\;$-13.877 871 0              & \cite{WZC2}     & 1424.3   \\
9  & 4$^2D$ & 646 & -13.778 890 & 1056 & -13.779 084 &   841 & $\;$-13.780 514 4              & \cite{WZC2}     & 1430.8   \\
14 & 5$^2D$ & 646 & -13.733 841 & 1056 & -13.734 024 &   841 & $\;$-13.735 455 4              & \cite{WZC2}     & 1431.8   \\
20 & 6$^2D$ & 646 & -13.709 377 & 1056 & -13.709 538 &       &                            &                 &          \\
27 & 7$^2D$ &     &             & 1056 & -13.694 804 &       &                            &                 &          \\
\hline
8  & 4$^2F$ & 286 & -13.779 403 &  532 & -13.779 946 &       &                            &                 &           \\
13 & 5$^2F$ & 286 & -13.734 363 &  532 & -13.734 924 &       &                            &                 &           \\
18 & 6$^2F$ & 286 & -13.709 788 &  532 & -13.710 457 &       &                            &                 &           \\
25 & 7$^2F$ &     &             &  532 & -13.695 579 &       &                            &                 &           \\
\hline
   &        &     &    n=7      &     &     n=8     &        &                            &                 &        \\   
12 & 5$^2G$ & 395 & -13.734 819 & 694 & -13.735 021 &        &                            &                 &        \\
17 & 6$^2G$ & 395 & -13.710 358 & 694 & -13.710 575 &        &                            &                 &        \\
24 & 7$^2G$ & 395 &             & 694 & -13.695 806 &        &                            &                 &        \\
\hline
16 & 6$^2H$ & 272 & -13.710 376 & 519 & -13.710 578 &        &                            &                 &        \\
23 & 7$^2H$ & 272 & -13.695 613 & 519 & -13.695 828 &        &                            &                 &        \\
\hline
22 & 7$^2I$ &     &             & 350 & -13.695 844 &        &                            &                 &        \\
\hline\hline
\end{tabular}}
\end{center}
\end{table}

\newpage


\begin{table}[tp]
\begin{center}
\caption{Calculated Hy-CI energies of the ground S-state and first S-, P-, and D-excited states of Li atom.
Convergence of the calculations and comparison with energy values of the literature are shown. $n$ is the odering number of the states. 
N is the number of symmetry adapted configurations. 
All energies are given in a.u., while Diff. are the energy differences between the present and reference energies in microhartrees ($1 \cdot 
10^{-6}$ a.u.).} 
\scalebox{0.90}{%
\begin{tabular}{rcccrcrlcc}
\hline\hline                                                              
No.  \quad & \quad  State \quad & \quad N \qquad & \qquad  n=4 \qquad & \qquad N \quad \qquad  & \qquad n=5-7 \qquad & \qquad N \qquad & 
\qquad  Ref. Ener. \quad & \quad Ref. \quad & \quad  Diff. \\ 
\hline 
 1&2$^2S$& 309 & -7.478 053 222 &  693 & -7.478 058 969 & 34020 & -7.478 060 323 910 146 894 & \cite{YD}      & 1.3 \\
 3&3$^2S$& 307 & -7.354 078 275 &  549 & -7.354 093 706 & 34020 & -7.354 098 421 444 364 045 & \cite{YD}      & 4.7 \\         
 6&4$^2S$& 252 & -7.318 481 008 &  591 & -7.318 517 759 & 34020 & -7.318 530 845 998 906 901 & \cite{YD}      & 13.  \\
10&5$^2S$&     &                &  687 & -7.303 496 699 & 34020 & -7.303 551 579 226 734 650 & \cite{YD}      & 54.9 \\
15&6$^2S$&     &                &  491 & -7.295 739 603 & 34020 & -7.295 859 510 844 131 039 & \cite{YD}      & 120.0\\ 
  &      &     &                &      &      n=8       &       &                            &                &      \\ 
24&7$^2S$&     &                &  506 & -7.291 085 910 & 17072 & -7.291 392 273 116         & \cite{Sims-Li} & 306.2 \\ 
28&8$^2S$&     &                &  506 & -7.288 321 853 &       &                            &                &      \\ 
\hline
 2&2$^2P$& 381 & -7.410 134 123 &  616 & -7.410 149 407 & 32200 & -7.410 156 532 652 370     & \cite{YD}    & 7.1 \\
 4&3$^2P$& 530 & -7.337 055 167 &  766 & -7.337 113 796 &  7000 & -7.337 151 707 93          & \cite{AdamP} & 37.9\\
 7&4$^2P$& 466 & -7.311 724 861 &  752 & -7.311 811 529 &  7000 & -7.311 889 059 38          & \cite{AdamP} & 77.5  \\
13&5$^2P$&     &                &  750 & -7.300 137 068 &  7000 & -7.300 288 164 88          & \cite{AdamP} & 151.1\\         
19&6$^2P$&     &                &  847 & -7.293 967 122 &  7000 & -7.294 020 052 93          & \cite{AdamP} & 52.9 \\
  &      &     &                &      &      n=8       &       &                            &                &      \\
26&7$^2P$&     &                &  502 & -7.289 814 402 &       &                            &              &      \\  
\hline
 5&3$^2D$& 188 & -7.335 505 135 &  490 & -7.335 512 623 & 32760 & -7.335 523 543 524 685    & \cite{YD}     &  10.9 \\         
 8&4$^2D$& 176 & -7.311 192 543 &  187 & -7.311 211 047 &  4000 & -7.311 189 578 43         & \cite{AdamD}  & -21.5 \\    
14&5$^2D$& 273 & -7.298 186 482 &  448 & -7.299 848 265 &  4000 & -7.299 927 555 94         & \cite{AdamD}  &  79.3 \\  
20&6$^2D$&     &                &  271 & -7.293 697 654 &  4000 & -7.293 810 713 64         & \cite{AdamD}  & 113.1 \\  
  &      &     &                &      &     n=8        &       &                           &               &       \\ 
27&7$^2D$&     &                &  423 & -7.289 806 792 &  4000 & -7.290 122 856 24         & \cite{AdamD}  & 324.1 \\    
\hline\hline
\end{tabular}}
\end{center}
\end{table}

\newpage


\begin{table}[tp]
\begin{center}
\caption{Calculated Hy-CI energies of the ground S-state and first S-, P-, and D-excited states of the Be$^+$ ion.  
Convergence of the calculations and comparison with energy values of the literature are shown here. $n$ is the ordering number of the states. 
N is the number of symmetry adapted configurations. All energies are given in a.u., while Diff. are the energy differences in microhartrees 
($1 \cdot 10^{-6}$ a.u.).}  
\scalebox{0.90}{%
\begin{tabular}{rcccrcrlcc}
\hline\hline
No. \quad & \quad  State \quad & \quad N \qquad & \qquad  n=4 \qquad & \qquad N \quad \qquad  & \qquad n=5,7 \qquad & \qquad N \qquad & 
\qquad  Ref. Ener. \quad & \quad Ref. \quad & \quad  Diff. \\
\hline                                                                                                                      
 1 & 2$^2S$ & 514 & -14.324 757 377 & 1028 & -14.324 761 678 & 13944 &  -14.324 763 176 790 150 & \cite{PKP}      &   1.5 \\      
 3 & 3$^2S$ & 502 & -13.922 759 980 & 1199 & -13.922 784 968 &~10000 &  -13.922 789 268 554 2   & \cite{PP}       &   4.3 \\      
 6 & 4$^2S$ & 409 & -13.798 520 453 &  757 & -13.798 706 849 & 1888  &  -13.798 716 609 2       & \cite{Adam-Be+} &   9.8 \\      
10 & 5$^2S$ &     &                 &  698 & -13.744 580 355 & 1940  &  -13.744 631 82          & \cite{King-Be+} &  51.5 \\    
14 & 6$^2S$ &     &                 &  560 & -13.716 205 613 & 2058  &  -13.716 286 24          & \cite{King-Be+} &  84.6 \\    
18 & 7$^2S$ &     &                 &  810 & -13.699 224 475 &       &                          &                 &      \\ 
   &        &     &                 &      &      n=8        &       &                          &                 &     \\
28 & 8$^2S$ &     &                 &  556 & -13.687 885 004 &       &                          &                 &       \\  
\hline
 2 & 2$^2P$ & 373 & -14.179 314 875 &  616 & -14.179 327 999 &~10000 &  -14.179 333 293 342 7   & \cite{PP}       &   5.3 \\   
 4 & 3$^2P$ & 499 & -13.885 035 680 &  707 & -13.885 115 345 &       &  -13.885 15              & \cite{Woznicki} &  34.7 \\  
 7 & 4$^2P$ & 352 & -13.783 432 326 &  582 & -13.783 574 124 &  1021 &  -13.783 518 3           & \cite{WZC1}     & -56.1 \\ 
11 & 5$^2P$ &     &                 &  674 & -13.736 485 889 &       &  -13.737 18              & \cite{Woznicki} & 694.0 \\
16 & 6$^2P$ &     &                 & 1232 & -13.711 935 268 &       &  -13.712 06              & \cite{Woznicki} & 124.7 \\
   &        &     &                 &      &      n=8        &       &                          &                 &     \\
20 & 7$^2P$ &     &                 &  503 & -13.696 356 527 &       &                          &                 &     \\
\hline
 5 & 3$^2D$ & 265 & -13.878 005 890 &  426 & -13.878 041 021 &  841  &  -13.877 871 0           & \cite{WZC2}     &-170.0 \\    
 9 & 4$^2D$ & 230 & -13.779 724 788 &  450 & -13.780 663 883 &  841  &  -13.780 514 4           & \cite{WZC2}     &-149.4 \\
13 & 5$^2D$ & 250 & -13.728 658 182 &  407 & -13.735 537 780 &  841  &  -13.735 455 4           & \cite{WZC2}     & -82.3 \\
17 & 6$^2D$ &     &                 &  444 & -13.710 204 495 &       &                          &                 &     \\       
   &        &     &                 &      &      n=8        &       &                          &                 &     \\ 
27 & 7$^2D$ &     &                 &  529 & -13.695 419 936 &       &                          &                 &      \\
\hline\hline                               
\end{tabular}}
\end{center}
\end{table}

\newpage


\begin{table}[tp]
\caption{The total energies of the S-, P-, D-, F-, G-, H-, and I-states of the Li atom (in a.u.) 
ordered by their energy. The bound states lay below the ionization threshold of the Li$^+$ ion, 
which non-relativistic values is -7.27991 34126 69305 96491 810(15) a.u. \cite{Frolov-Li+}.}
\begin{center}
\scalebox{0.90}{
\begin{tabular}{rclllc}
\hline\hline
No. & \quad State \qquad & \qquad \quad  E(FCI) \quad \qquad & \quad \qquad E(Hy-CI) \qquad \quad & Ref. Energy & Ref. \\ 
\hline
 1 & 2$^2S$ & -7.477 20(1)  & -7.478 060(2)  & -7.478 060 323 910 147(1)  & \cite{YD} \\ 
 2 & 2$^2P$ & -7.408 70(9)  & -7.410 150(6)  & -7.410 156 532 652 41(4)   & \cite{YD} \\ 
 3 & 3$^2S$ & -7.353 25(1)  & -7.354 095(2)  & -7.354 098 421 444 37(1)   & \cite{YD} \\ 
 4 & 3$^2P$ & -7.335 70(4)  & -7.337 120(7)  & -7.337 151 707 93          & \cite{AdamP} \\
 5 & 3$^2D$ & -7.334 20(9)  & -7.335 520(8)  & -7.335 523 543 524 688(3)  & \cite{YD} \\ 
 6 & 4$^2S$ & -7.317 70(3)  & -7.318 520(3)  & -7.318 530 845 998 91(1)   & \cite{YD} \\ 
 7 & 4$^2P$ & -7.310 40(2)  & -7.311 820(9)  & -7.311 889 059 38          & \cite{AdamP} \\
 8 & 4$^2D$ & -7.309 80(4)  & -7.311 220(9)  & -7.311 189 578 43(200)     & \cite{AdamD} \\ 
 9 & 4$^2F$ & -7.309 60(9)  &  &  &  \\ 
10 & 5$^2S$ & -7.302 70(2)  & -7.303 50(4)   & -7.303 551 579 226 77(4)   & \cite{YD} \\ 
11 & 5$^2G$ & -7.299 50(7)  &  &  &  \\   
12 & 5$^2F$ & -7.299 40(6)  &  &  &  \\ 
13 & 5$^2P$ & -7.298 90(10) & -7.300 20(7)   & -7.300 288 164 88          & \cite{AdamP} \\
14 & 5$^2D$ & -7.298 60(10) & -7.299 90(5)   & -7.299 927 555 94(300)     & \cite{AdamD} \\
15 & 6$^2S$ & -7.296 00(6)  & -7.295 80(6)   & -7.295 859 510 844 19(6)   & \cite{YD}  \\ 
16 & 6$^2H$ & -7.293 40(8)  &  &  &  \\
17 & 6$^2G$ & -7.293 30(1)  &  &  &  \\
18 & 6$^2F$ & -7.293 30(9)  &  &  &  \\ 
19 & 6$^2P$ & -7.293 60(5)  & -7.294 00(4)   & -7.294 020 052 93          & \cite{AdamP} \\
20 & 6$^2D$ & -7.292 40(1)  & -7.293 70(3)   & -7.293 810 713 64(500)     & \cite{AdamD} \\
21 & 7$^2I$ & -7.289 71(7)  &  &  &  \\ 
22 & 7$^2H$ & -7.289 70(7)  &  &  &  \\
23 & 7$^2G$ & -7.289 65(4)  &  &  &  \\ 
24 & 7$^2S$ & -7.285 70(5)  & -7.291 10(15)  & -7.291 392 276(3)          & \cite{Sims-Li}  \\ 
25 & 7$^2F$ & -7.289 45(5)  &  &  &  \\ 
26 & 7$^2P$ & -7.289 8(5)   & -7.290 00(9)   &  &  \\ 
27 & 7$^2D$ & -7.289 8(10)  & -7.289 90(10)  & -7.290 122 856 24(2000)    & \cite{AdamD} \\ 
28 & 8$^2S$ & -7.286 7(10)  & -7.288 50(18)  &  &  \\ 
\hline\hline
\end{tabular}
}
\end{center}
\end{table}

\newpage


\begin{table}[tp]
\caption{The total energies of the S-, P-, D-, F-, G-, H-, and I-states of the Be$^+$ ion ordered by their total energies. 
All these bound states are stable, since their total energies are below the ionization threshold which
coincide with the non-relativistic energy of the Be$^{2+}$ ion $E_{tr}$ = -13.65556 62384 23586 70207 810(15)
a.u. \cite{Frolov-Li+}.} 
\begin{center}
\scalebox{0.90}{
\begin{tabular}{rclllc}
\hline\hline
No.& \quad State \qquad & \qquad \quad  E(FCI) \quad \qquad & \quad \qquad E(Hy-CI) \qquad \quad & Ref. Energy & Ref. \\ \hline
 1 & 2$^2S$ & -14.323 80(3)  & -14.324 763(2)  & -14.324 763 176 790 43(22) & \cite{PKP} \\ 
 2 & 2$^2P$ & -14.177 43(2)  & -14.179 330(3)  & -14.179 333 293 42(3)   & \cite{PP} \\ 
 3 & 3$^2S$ & -13.921 90(3)  & -13.922 786(2)  & -13.922 789 268 570(10) & \cite{PP} \\ 
 4 & 3$^2P$ & -13.883 50(7)  & -13.885 120(5)  & -13.885 15              & \cite{Woznicki} \\        
 5 & 3$^2D$ & -13.876 50(5)  & -13.878 050(5)  & -13.877 871 0           & \cite{WZC2} \\ 
 6 & 4$^2S$ & -13.797 80(5)  & -13.798 710(4)  & -13.798 716 609 2       & \cite{Adam-Be+} \\    
 7 & 4$^2P$ & -13.782 00(2)  & -13.783 580(5)  & -13.783 518 3           & \cite{WZC1} \\    
 8 & 4$^2F$ & -13.778 00(5)  &  &  &  \\ 
 9 & 4$^2D$ & -13.779 10(2)  & -13.780 70(4)   & -13.780 514 4           & \cite{WZC2} \\ 
10 & 5$^2S$ & -13.743 70(4)  & -13.744 60(2)   & -13.744 631 82          & \cite{King-Be+} \\
11 & 5$^2P$ & -13.735 50(3)  & -13.736 50(2)   & -13.737 18              & \cite{Woznicki} \\    
12 & 5$^2G$ & -13.735 10(8)  &  &  &  \\ 
13 & 5$^2F$ & -13.735 00(7)  &  &  &  \\ 
14 & 5$^2D$ & -13.734 10(7)  & -13.735 60(6)   & -13.735 455 4           & \cite{WZC2} \\ 
15 & 6$^2S$ & -13.715 30(8)  & -13.716 25(5)   & -13.716 286 24          & \cite{King-Be+} \\ 
16 & 6$^2H$ & -13.710 65(2)  &  &  &  \\ 
17 & 6$^2G$ & -13.710 60(3)  &  &  &  \\ 
18 & 6$^2F$ & -13.710 50(4)  &  &  &  \\ 
19 & 6$^2P$ & -13.710 20(6)  & -13.712 00(7)   & -13.712 06              & \cite{Woznicki} \\    
20 & 6$^2D$ & -13.709 60(6)  & -13.710 25(5)   &  &  \\ 
21 & 7$^2S$ & -13.697 50(8)  & -13.699 30(8)   &  &  \\
22 & 7$^2I$ & -13.695 86(2)  &  &  &  \\  
23 & 7$^2H$ & -13.695 85(2)  &  &  &  \\
24 & 7$^2G$ & -13.695 83(2)  &  &  &  \\
25 & 7$^2F$ & -13.695 60(2)  &  &  &  \\ 
26 & 7$^2P$ & -13.695 3(7)   & -13.696 40(5)   &  &  \\ 
27 & 7$^2D$ & -13.694 9(10)  & -13.695 50(8)   &  &  \\ 
28 & 8$^2S$ & -13.684 8(3)   & -13.688 00(12)  &  &  \\ 
\hline\hline
\end{tabular}
}
\end{center}
\end{table}

\end{document}